\documentstyle[12pt,epsf]{article}\textheight 230mm\textwidth 160mm
\newcommand{\bi}{\bibitem}
\newcommand{\be}{\begin{eqnarray}}
\newcommand{\ee}{\end{eqnarray}} 
\newcommand{\nn}{\nonumber}
\catcode`\@=11
\def\lsim{\mathrel{\mathpalette\@versim<}}
\def\gsim{\mathrel{\mathpalette\@versim>}}
\def\@versim#1#2{\vcenter{\offinterlineskip
\ialign{$\m@th#1\hfil##\hfil$\crcr#2\crcr\sim\crcr } }}
\catcode`\@=12
\begin{document}
\hspace*{11cm}\vspace{-3mm}MPI-Ph/95-125\\ 
\hspace*{11.7cm}\vspace{-3mm}KANAZAWA-95-17\\
\hspace*{11.7cm}November 1995
\begin{center}
{\Large\bf  Gauge-Yukawa Unification\\
in \\$SO(10)$ SUSY GUTs}
\end{center} 

\vspace{0.5cm}

\begin{center}{\sc Jisuke Kubo}$\ ^{1,*}$, \vspace{-1mm}
{\sc Myriam Mondrag{\' o}n}$\ ^{2,**}$, \vspace{-1mm}\\
{\sc Shigeyuki Shoda}$\ ^{3}$ \vspace{-1mm}and 
{\sc George Zoupanos}$\ ^{4,***}$  
\end{center}
\begin{center}
{\em $\ ^{(1)}$ 
 College of Liberal Arts, Kanazawa \vspace{-2mm} University,
Kanazawa 920-11, Japan } \\
{\em $\ ^{(2)}$ Institut f{\" u}r Theoretische Physik,
Philosophenweg 16 \vspace{-2mm}\\
D-69120 Heidelberg, Germany}\vspace{-2mm}\\
{\em $\ ^{(3)}$ 
 Physics Department, Kanazawa \vspace{-2mm} University,
Kanazawa 920-11, Japan }\\ 
{\em $\ ^{(4)}$ Max-Planck-Institut f\"ur Physik,
 Werner-Heisenberg-Institut \vspace{-2mm}\\
D-80805 Munich, Germany} 
\end{center}

\noindent
{\sc\large Abstract}
\newline
\noindent
We study
supersymmetric unified models 
with  three fermion generations
 based on the gauge group
 $SO(10)$  and require 
 Gauge-Yukawa Unification, i.e.,
a renormalization group invariant 
functional relationship among the gauge and Yukawa couplings of 
the third generation in the symmetric phase.
In the case of the minimal model,
 we find that the predicted values for
 the top and  bottom quark masses 
are in agreement with the present experimental data
for a wide range of supersymmetry breaking scales.
We also find that an experimental accuracy of
less than $1 \%$ for the top quark mass  
could 
test the corresponding  prediction of the Gauge-Yukawa unified model.
\vspace*{1cm}
\footnoterule
\noindent
$^{*}$Partially supported \vspace{-2mm} by the Grants-in-Aid
for \vspace{-2mm} Scientific Research  from the Ministry of
Education, Science  \vspace{-2mm}
and Culture (No. 40211213).\\
\noindent
$^{**}$Address \vspace{-2mm} after Jan. 1st, '96:
Instituto de F{\' \i}sica, \vspace{-2mm} UNAM,
Apdo. Postal 20-364,
M{\' e}xico 01000 D.F., \vspace{-2mm} M{\' e}xico.\\
\noindent
$^{***}$ On leave of absence from \vspace{-2mm}
 Physics Department, National
Technical\vspace{-2mm} University, GR-157 80 Zografou, Athens,
\vspace{-2mm} Greece.\\  
Partially supported by \vspace{-2mm} C.E.C. projects, SC1-CT91-0729;
CHRX-CT93-0319.
\newpage
\pagestyle{plain}

\section{Introduction} 

\vspace{0.5cm}
\noindent
The remarkable success of the standard model (SM)  suggests
that we have at hand a highly non-trivial part of a more fundamental
theory for elementary particle physics. 
Since, however, the SM contains many independent parameters,
it has been a challenge  to understand the plethora
 of these free parameters. 
In Grand Unified Theories (GUTs) \cite{georgi1,fritzsch1}, 
 the gauge interactions of the SM are unified at a 
certain energy scale $M_{\rm GUT}$, and consequently its gauge
couplings are related with each other. Also the  Yukawa couplings
can be related among themselves to a certain extent.
These relations among the couplings can yield
 testable predictions  for  GUTs \cite{georgi2,buras1}.

 However, GUTs
can not relate the gauge and
Yukawa couplings with each other. In order to achieve Gauge-Yukawa 
Unification (GYU), within the assumption that all the particles
appearing in a field theory model are elementary, one has to
consider extended supersymmetry \cite{fayet1}.
Unfortunately,
it is extremely difficult to construct
a realistic model based on the extended supersymmetry,
because the model has a real structure with respect
to $SU(2)_{L}\times U(1)_{Y}$ \cite{fayet1}.
In superstrings and composite models such relations, 
in principle, also
exist.
However, in both cases
 there exist open difficult problems which  among others are 
 related to 
the lack of realistic models.

Recently, an alternative way to achieve
unification of couplings 
\cite{zimmermann1}-\cite{kubo4} has been proposed;
it is based on the
fact that within the framework of 
a renormalizable field theory, one can find renormalization group (RG)
invariant relations among parameters which
can improve the calculability and the predictive power
of the theory.
This idea is called sometimes the principle 
of reduction couplings \cite{zimmermann1}.
In this paper we would like to consider $SO(10)$ 
supersymmetric GUTs along the lines of this
unification idea.
We note that all realistic supersymmetric $SO(10)$ models have to be
asymptotically-nonfree,
because one needs a certain set of Higgses to break
$SO(10)$ down to $SU(3)_{C}\times SU(2)_L \times U(1)_Y $
\cite{fritzsch1,mohapatra1}. 

The common wisdom is that the
asymptotically-nonfree theories develop a Landau pole at a high energy
scale, a fact which inevitably suggests that the theory is trivial,
unless new physics is entering before the couplings blow up.
However, there exist arguments
leading to a different view point;
the theory  converges to
a well-defined ultraviolet fixed  point, a ''new phase'',
 instead  of blowing up  \footnote{See 
ref.  \cite{kondo1}, which contain also earlier references on
ultraviolet fixed points.}.  
Non-abelian gauge theories 
could have the same behavior,
and it might be that
an asymptotically nonfree, non-abelian gauge theory
with matter couplings can change  after the
critical value of the couplings its phase due to a certain 
self-adjustment of the couplings and become 
a well defined finite theory \footnote{Similar
phenomenon has been observed in 
asymptotically free theories,
in which the couplings have to be related
with each other in order for the theories to be
asymptotically free and hence well defined in
the ultraviolet limit \cite{kubo2,harada1}.}.
In this way, a dynamical unification 
of couplings  \cite{kubo5} \footnote{This analysis
was motivated by the observation  \cite{st1} that
 there might exist
an infrared fixed point in asymptotically free QCD.} 
 can be achieved, since to enter into the new phase the
couplings are supposed to satisfy  a definite relation.
It is natural to assume that at scales below the critical value,
that is, in the symmetric phase of a GUT
these relations among the couplings are RG invariant as
a remnant of the dynamical unification of couplings.
When the GUT enters in its spontaneous broken phase,
these RG invariant relations serve just as boundary
conditions at $M_{\rm GUT}$ on the evolution of 
couplings for  scales below it.
Since the principle of reduction of couplings
is based on RG invariant relations among
couplings, the GYU based on this principle
could be a consequence of the dynamical unification
of couplings described above.
This is a speculation, of course,
because so far there exists no reliable
and decisive  calculation
on the behavior of asymptotically-nonfree,
non-abelian gauge theories \footnote{Even the case of 
QED  has not been 
completely clarified  \cite{kogut1,gockeler1}.}.

There have been recently various phenomenological studies on
$SO(10)$ supersymmetric GUTs without GYU \cite{so10}-\cite{so101},
where the top quark mass was calculated from the requirement
of the correct bottom-tau hierarchy
(recall that the top Yukawa coupling contributes 
significantly to the
RG evolution  of the bottom  and tau Yukawa couplings).
Unfortunately, there exists a wide range of the predicted values 
of $M_t$ ($160-200$ GeV \cite{so102}). 
One of the main reasons is that
 the  bottom-tau hierarchy is experimentally known
only with a large uncertainty ($\sim 10 \%$ \cite{pdg}).
Suppose that the  bottom-tau hierarchy would
be precisely known and the calculated top mass would 
exactly agree with the experimental value. Then there should exist
a unique Yukawa coupling of the third generation 
in $SO(10)$ GUTs, which is consistent with 
experimental data.
It is, however, clear  that this Yukawa coupling can not be predicted
within the conventional GUT scheme.

With  a GYU the model obtains a more predictive power, and the 
Yukawa couplings become calculable, as has been experienced
in our recent studies on other models \cite{kapet1}-\cite{kubo4}.
Although the GYU proposed there is a gradual, conservative
extension of the usual GUT scheme, it has turned out to yield
successful predictions.
We emphasize that this success
is not just a consequence of the infrared behavior of the Yukawa
couplings \cite{infra}.
 Although the infrared behavior  is an important ingredient for the
successful predictions,  we should stress also
the significance of the field content of the theories as well as
their interactions above the unification scale.
 The reason is that they contribute to the
$\beta$ functions in the symmetric phase
 which fix the structure of the GYU based on the principle
of reduction of couplings and consequently
the boundary conditions  for the evolution of couplings
below the GUT scale.
Therefore, it is absolutely nontrivial that
(1) there exist a unique Yukawa coupling of the third generation
in a supersymmetric $SO(10)$ GUT that
is consistent with the experimental data, and (2) 
this Yukawa coupling can be calculated by means of the reduction
of couplings.

In this paper we will consider two different models;
the first one with
the Higgs supermultiplets of 
${\bf 1},{\bf 10},{\bf 16},\overline{{\bf 16}},
{\bf 45},{\bf 54}$ and 
 the three fermion generations in ${\bf 16}$, 
and the second one with ${\bf 126}+\overline{\bf 126}$
instead of the singlet which provides the Georgi-Jarlskog
mechanism \cite{gj,mohapatra1} to obtain a realistic mass fermion
matrix. We have found that the first model yields 
experimentally consistent predictions, where we neglect
the Yukawa coulings of the first two fermion generations.
For the second model we obtain 
couplings which are so large that the model either cannot be treated
within the framework of perturbation theory or
does not give a testable prediction on the top-bottom hierarchy.

Our approach to predict the top and
bottom quark masses, 
at first sight, looks similar to
the infrared-fixed-point approach of ref. \cite{ross1}.
In the final section of this paper, we will discuss
this approach within the framework of our concrete 
$SO(10)$ model and conclude
that the infrared-fixed-point approach  
does not always provide us with precise
predictions on low energy parameters.

\section{The models} 

We denote the hermitean
$SO(10)$-gamma matrices  by
$\Gamma_{\alpha}~,~\alpha=1,\cdots,10$.
The charge conjugation matrix $C$ satisfies
$C = C^{-1}~,~C^{-1}\,\Gamma_{\alpha}^{T}\,C ~=~
-\Gamma_{\alpha}$, and 
the $\Gamma_{11}$ is defined as
$\Gamma_{11} \equiv (-i)^5 \,\Pi_{\alpha =1}^{10}
\Gamma_{\alpha} ~~ \mbox{with} ~~(\Gamma_{11})^2 ~=~1$.
The chiral projection operators are given by
${\cal P}_{\pm} = \frac{1}{2}(\,1\pm \Gamma_{11})$.

In $SO(10)$ GUTs \cite{fritzsch1,mohapatra1},
three generations of quarks and leptons are
 accommodated by 
three chiral supermultiplets in
${\bf 16}$ which we
denote by 
\be
\Psi^{I}({\bf 16})~~\mbox{with}~~{\cal P}_{+}\,
\Psi^{I}~=~\Psi^{I}~,
\ee
where $I$ runs over the three generations
and the spinor index is suppressed.
To break $SO(10)$ down to $SU(3)_{\rm C}
\times SU(2)_{\rm L} \times U(1)_{\rm Y}$, we use
the following set of chiral superfields:
\be
S_{\{\alpha\beta\}}({\bf 54})~,~
A_{[\alpha\beta]}({\bf 45})~,~
\phi({\bf 16})~,~\overline{\phi}({\overline{\bf 16}})~.
\ee
The two $SU(2)_{\rm L}$ doublets which are responsible for
the spontaneous symmetry breaking (SSB) of 
$SU(2)_{\rm L} \times U(1)_{\rm Y}$ down to $U(1)_{\rm EM}$
are contained in
$ H_{\alpha}({\bf 10})$.
\subsection{Model I}
For model I, we further introduce a singlet $\varphi$ which after
the  SSB of $SO(10)$ will
mix with the right-handed neutrinos 
so that they will become superheavy.

The superpotential of  model I is 
given by 
\be
W^{I} &=& W_{Y}+W_{SB}+ W_{HS}+ W_{NM}^{I}+W_{M}~,
\ee
where
\be
W_{Y} &=&\frac{1}{2}\sum_{I,J=1}^{3}g_{IJ}\,\Psi^{I}\,C\Gamma_{\alpha} 
\,\Psi^{J}~H_{\alpha}~,\nn\\
W_{SB} &=&\frac{g_{\phi}}{2}
\,\overline{\phi}\,
\Gamma_{[\alpha\beta]}\,\phi~A_{[\alpha\beta]}
+\frac{g_{S}}{3!}\,
\mbox{Tr}~S^3+
\frac{g_{A}}{2}\,\mbox{Tr}~A^2\,S~, \\
W_{HS} &=&
\frac{g_{HS}}{2}\,\
H_{\alpha}\,S_{ \{\alpha\beta \} }\,H_{\beta}~,~
W_{NM}^{I} ~=~ \sum_{I=1}^{3}\,g_{INM}\,\Psi^I\,
\overline{\phi}\,\varphi~,\nn\\
W_{M}&=&\frac{m_{H}}{2}\,H^2+m_{\varphi}\,\varphi^2
+m_{\phi}\,\overline{\phi} \phi+\frac{m_{S}}{2}\,S^2+
\frac{m_{A}}{2}\,A^2~ ,\nn 
\ee
and $\Gamma_{[\alpha\beta]}=i
 (\Gamma_{\alpha}\Gamma_{\beta}-
\Gamma_{\beta}\Gamma_{\alpha})/2$.
The superpotential is not the most general one, but
by virtue of the non-renormalization theorem,
this does not contradict the philosophy of 
the coupling unification by the reduction 
method (a RG invariant fine tuning is a solution
of the reduction equation) .
$W_{SB}$  is responsible for
the SSB of $SO(10)$ down to
$SU(3)_{C}\times SU(2)_{W}\times U(1)_{Y}$, 
and this can be achieved without breaking supersymmetry,
while $W_{HS}$ is responsible for
the triplet-doublet splitting of $H$.
The right-handed neutrinos 
obtain a superheavy mass through $W_{NM}$ after the
SSB (as announced), and the Yukawa couplings 
for the leptons and quarks are contained in $W_{Y}$.
We assume  that
 there exists a choice of soft supersymmetry breaking terms so that
all the vacuum expectation values necessary for the desired SSB 
corresponds to the minimum of
the potential.

Given the supermultiplet content and the superpotential $W$,
we  can  compute the $\beta$ functions of the model. 
The gauge coupling of $SO(10)$ is denoted by $g$, and 
our  normalization of the $\beta$ functions
is as usual, i.e., 
$d g_{i}/d \ln \mu ~=~
\beta^{(1)}_{i}/16 \pi^2+O(g^5)$,
where $\mu$ is the renormalization
scale. We find: 
\be
  \beta^{(1)}_{g} &=&7\,g^3~,\nn\\
  \beta^{(1)}_{g_T} &=& g_T\,(\,14 |g_T|^2+\frac{27}{5}|g_{HS}|^2+
|g_{3NM}|^2-\frac{63}{2}g^2\,)~,\nn\\
  \beta^{(1)}_{g_{\phi}} &=& g_{\phi}(\,53 |g_{\phi|^2}+
\frac{48}{5}|g_{A}|^2+\frac{1}{2}|g_{1NM}|^2+\frac{1}{2}|g_{2NM}|^2+
\frac{1}{2}|g_{3NM}|^2-\frac{77}{2}g^2\,)~,\nn\\
\beta^{(1)}_{S} &=&g_{S}(\,\frac{84}{5}|g_{S}|^2
+12|g_{A}|^2+\frac{3}{2}|g_{HS}|^2-60 g^2\,)~,\nn\\
\beta^{(1)}_{A} &=&
g_{A}(\,16|g_{\phi}|^2+
\frac{28}{5}|g_{S}|^2+\frac{116}{5}|g_{A}|^2+\frac{1}{2}|g_{HS}|^2 -52
g^2\,)~,\nn\\
  \beta^{(1)}_{HS} &=&g_{HS}(\,8|g_{T}|^2+
\frac{28}{5}|g_{S}|^2+4|g_{A}|^2+\frac{113}{10}|g_{HS}|^2
-38 g^2\,) ~,\\
\beta^{(1)}_{1NM} &=& g_{1NM}(\,\frac{45}{2}|g_{\phi}|^2+
9 |g_{1NM}|^2+\frac{17}{2}|g_{2NM}|^2+
\frac{17}{2}|g_{3NM}|^2-\frac{45}{2}g^2\,)~,\nn\\
 \beta^{(1)}_{2NM} &=& g_{2NM}(\,\frac{45}{2}|g_{\phi}|^2+
\frac{17}{2} |g_{1NM}|^2+9 |g_{2NM}|^2+
\frac{17}{2}|g_{3NM}|^2-\frac{45}{2}g^2\,)~,\nn\\
\beta^{(1)}_{3NM} &=& g_{3NM}(\,5 |g_T|^2+\frac{45}{2}|g_{\phi}|^2+
\frac{17}{2} |g_{1NM}|^2+\frac{17}{2}|g_{2NM}|^2+
9|g_{3NM}|^2-\frac{45}{2}g^2\,)~.\nn
\ee
We have assumed that the Yukawa couplings $g_{IJ}$ except for
$g_T \equiv g_{33}$ vanish. They can be included as small
perturbations \footnote{We will clarify later what we mean
by small perturbations.}. Needless to say that the
soft susy breaking terms do not alter the $\beta$ functions
above.
\subsection{Model II}
For model II, we introduce a pair of
\be
\Theta_{[\alpha\beta\gamma\mu\nu]}({\bf 126})~~\mbox{and}~~
\overline{\Theta}_{
[\alpha\beta\gamma\mu\nu]}(\overline{{\bf 126}})~
\ee
instead of the singlet $\varphi$,  providing us with
a possibility of incorporating the 
Georgi-Jarlskog mechanism \cite{gj,mohapatra1}.
They satisfy the duality conditions
\be
\Theta_{[\alpha_1 \cdots \alpha_{5}]}~
(\overline{\Theta}_{[\alpha_1 \cdots \alpha_{5}]}) &=&
-(+)\frac{i}{5!}\epsilon_{\alpha_1 \cdots \alpha_{10}}\,
\Theta_{[\alpha_6 \cdots \alpha_{10}]}~
(\overline{\Theta}_{[\alpha_6 \cdots \alpha_{10}]})~,
\ee
and $\overline{\Theta}$ (instead of $\varphi$)
 also will mix with the right-handed
neutrinos to make them superheavy.

The superpotential of  model II is 
given by 
\be
W^{II} &=& W_{Y} + W_{SB}+ W_{HS} +W_{\Theta}+W_{GJ}+ W_{NM}^{II} 
+W_{M}~,
\ee
where
\be
W_{\Theta} &=&
\frac{g_{\Theta S}}{4! 8}\,
\Theta_{[\alpha_1\cdots \alpha_5]}\,
S_{ \{\alpha_5\beta_1 \} }\,\Theta_{[\beta_1\cdots\beta_5]}+
\frac{g_{\overline{\Theta} S}}{4! 8}\,
\overline{\Theta}_{[\alpha_1\cdots \alpha_5]}\,
S_{ \{ \alpha_5\beta_1\} }\,
\overline{\Theta}_{[\beta_1\cdots\beta_5]}\nn\\
& &+\frac{g_{\Theta A}}{4! 4}\,
\overline{\Theta}_{[\alpha_1\cdots \alpha_5]}\,
A_{[\alpha_5\beta_1]}\,\Theta_{[\beta_1\cdots\beta_5]}~,\nn\\
W_{GJ} &=&\frac{g_{GJ}}{5! 4}\,
\Psi^{2}\,C\Gamma_{[\alpha_1 \cdots \alpha_5]}\,\Psi^{2}~
\overline{\Theta}_{[\alpha_1 \cdots \alpha_5]}~,\nn\\
W_{NM}^{II} &=& \frac{1}{5!2}\,\sum_{I=1}^{3}\,g_{INM}\,\Psi^I\,
C\Gamma_{[\alpha_1 \cdots \alpha_5]}\,\phi~ 
\overline{\Theta}_{[\alpha_1 \cdots \alpha_5]}~, \nn
\ee
and $\Gamma_{[\alpha_1 \cdots \alpha_5]}=
\frac{1}{ 5!}\,(\,\Gamma_{\alpha_1}\cdots \Gamma_{\alpha_5}+~
\mbox{anti-symmetric permutations}~)$. 
($W_{Y}~,~W_{SB}~,~W_{HS}$ and $W_{M} $ are 
given in eq. (4), and  
$ W_M$  contains the $\Theta-\overline{\Theta}$ mass term instead of
the $\varphi$ mass term.) The $\beta$ functions are 
found to be
\be
  \beta^{(1)}_{g} &=&77\,g^3~,\nn\\
  \beta^{(1)}_{g_T} &=& g_T\,(\,14 |g_T|^2+\frac{27}{5}|g_{HS}|^2
+126|g_{3NM}|^2-\frac{63}{2}g^2\,)~,\nn\\
  \beta^{(1)}_{g_{\phi}} &=& g_{\phi}(\,53 |g_{\phi|^2}+
\frac{48}{5}|g_{A}|^2+63|g_{1NM}|^2+63|g_{2NM}|^2+
63|g_{3NM}|^2+\frac{35}{4}|g_{\Theta A}|^2
-\frac{77}{2}g^2\,)~,\nn\\
\beta^{(1)}_{S} &=&g_{S}(\,\frac{84}{5}|g_{S}|^2
+12|g_{A}|^2+\frac{3}{2}|g_{HS}|^2+\frac{105}{8}|g_{\Theta S}|^2+
\frac{105}{8}|g_{\overline{\Theta} S}|^2
-60 g^2\,)~,\nn\\
\beta^{(1)}_{A} &=&
g_{A}(\,16|g_{\phi}|^2+
\frac{28}{5}|g_{S}|^2+\frac{116}{5}|g_{A}|^2
+\frac{1}{2}|g_{HS}|^2 
 +\frac{35}{2}|g_{\Theta A}|^2+
\frac{35}{8}|g_{\overline{\Theta} S}|^2\nn\\
& &+\frac{35}{8}|g_{\Theta S}|^2-52 g^2\,)~,\nn\\
\beta^{(1)}_{HS} &=&g_{HS}(\,8|g_{T}|^2+
\frac{28}{5}|g_{S}|^2+4|g_{A}|^2+\frac{113}{10}|g_{HS}|^2
+\frac{35}{8}|g_{\overline{\Theta} S}|^2+
\frac{35}{8}|g_{\Theta S}|^2-38 g^2\,) ~,\\
\beta^{(1)}_{1NM} &=& g_{1NM}(\,\frac{45}{2}|g_{\phi}|^2+
134 |g_{1NM}|^2+71|g_{2NM}|^2+
71|g_{3NM}|^2
+4|g_{GJ}|^2\nn\\
& &+\frac{25}{8}|g_{\Theta A}|^2+
25|g_{\overline{\Theta} S}|^2
-\frac{95}{2}g^2\,)~,\nn\\
 \beta^{(1)}_{2NM} &=& g_{2NM}(\,\frac{45}{2}|g_{\phi}|^2+
71 |g_{1NM}|^2+134|g_{2NM}|^2+
71|g_{3NM}|^2+67|g_{GJ}|^2\nn\\
& &+\frac{25}{8}|g_{\Theta A}|^2+
25|g_{\overline{\Theta} S}|^2
-\frac{95}{2}g^2\,)~,\nn\\
\beta^{(1)}_{3NM} &=& g_{3NM}(\,5|g_{T}|^2+
\frac{45}{2}|g_{\phi}|^2+
71 |g_{1NM}|^2+71|g_{2NM}|^2+
134|g_{3NM}|^2\nn\\
& & +4|g_{GJ}|^2+
\frac{25}{8}|g_{\Theta A}|^2+
25|g_{\overline{\Theta} S}|^2
-\frac{95}{2}g^2\,)~,\nn\\
\beta^{(1)}_{\Theta S} &=&g_{\Theta S}(\,
\frac{28}{5}|g_{S}|^2+4|g_{A}|^2+\frac{1}{2}|g_{HS}|^2
+\frac{25}{4}|g_{\Theta A}|^2+
\frac{35}{8}|g_{\overline{\Theta} S}|^2\nn\\
& &+\frac{85}{8}|g_{\Theta S}|^2-70 g^2\,) ~,\nn\\
\beta^{(1)}_{\overline{\Theta} S} &=&
g_{\overline{\Theta} S}(\,\frac{28}{5}|g_{S}|^2+4|g_{A}|^2
+\frac{1}{2}|g_{HS}|^2
+8|g_{GJ}|^2+16|g_{1NM}|^2\nn\\
& &+16|g_{2NM}|^2
+16|g_{3NM}|^2+\frac{25}{4}|g_{\Theta A}|^2+
\frac{435}{8}|g_{\overline{\Theta} S}|^2
+\frac{35}{8}|g_{\Theta S}|^2-70 g^2\,)~,\nn\\
\beta^{(1)}_{\Theta A} &=& g_{\Theta A}(\,
\frac{48}{5}|g_{A}|^2
+8|g_{\phi}|^2
+4|g_{GJ}|^2+8|g_{1NM}|^2+8|g_{2NM}|^2
+8|g_{3NM}|^2\nn\\
& &+15|g_{\Theta A}|^2+
25|g_{\overline{\Theta} S}|^2
+\frac{25}{8}|g_{\Theta S}|^2-66 g^2\,)~,\nn\\
\beta^{(1)}_{GJ} &=&g_{GJ}(\,130|g_{GJ}|^2+
8|g_{1NM}|^2+134|g_{2NM}|^2+
8|g_{3NM}|^2+
\frac{25}{8}|g_{\Theta A}|^2+
25|g_{\overline{\Theta} S}|^2
-\frac{95}{2}g^2\,)~.\nn 
\ee
Observe the occurrence of large coefficients in the $\beta$ functions
above. They are responsible for the fact that 
the model II either cannot be treated in perturbation theory
or does not give a testable prediction on the top quark mass,
as we will see.
 
\section{Gauge-Yukawa Unification} 

The principle of reduction of coupling is to
impose as many as possible RG invariant constraints
which are compatible with renormalizability \cite{zimmermann1}.
Such constraints in the space of couplings can be expressed in the
implicit form  as $\Phi (g_1,\cdots,g_N) ~=~\mbox{const.}$, which
has to satisfy the partial differential equation
\be
{\vec \beta}\cdot {\vec \nabla}\,\Phi &=&\sum_{i=0}^{N}
\,\beta_{i}\,\frac{\partial}{\partial g_{i}}\,\Phi~=~0~,
\ee
where $\beta_i$ is the $\beta$ function of $g_i$.
In general,
 there exist,
at least locally, $N$ independent solutions of (10),
and they are equivalent to the solutions of
the so-called reduction equations \cite{zimmermann1},
\be
\beta \,\frac{d g_{i}}{d g} &=&\beta_{i}~,~i=1,\cdots,N~,
\ee
where $g\equiv g_0$ and 
$\beta\equiv\beta_0 $.
Since maximally $N$ independent 
RG invariant constraints
in the $(N+1)$-dimensional space of couplings
can be imposed by $\Phi_i$, one could in principle
express all the couplings in terms of 
a single coupling, the primary coupling $g$
\cite{zimmermann1}. This possibility is without any doubt
attractive, but  it can be unrealistic. Therefore, one often would
like to impose fewer RG invariant constraints,
leading to  the idea of
partial reduction \cite{kubo1,kubo2}. 

Here we would like to
briefly outline the method \footnote{
Detailed discussions on partial reduction are given in ref.
\cite{kubo3}, for instance.}.
For the case at hand, it is
convenient to work with the absolute square of $g_{i}$, and we
define the tilde couplings by 
\be
\tilde{\alpha}_{i} &\equiv& 
\frac{\alpha_{i}}{\alpha}~,~i=1,\cdots,N~,\nn
\ee
where 
$ \alpha ~=~|g|^2/4\pi$ and $\alpha_{i} ~=~ 
|g_{i}|^2/4\pi$.
We assume that their evolution equations take the form
\be
\frac{d \alpha}{dt} &=&-b^{(1)}\,\alpha^2+\cdots~,\nn\\
\frac{d\alpha_i}{dt} &=&-b^{(1)}_{i}\,\alpha_{i}\alpha+
\sum_{j,k}b^{(1)}_{i,jk}\,\,\alpha_j\alpha_k+\cdots~,
\ee
in perturbation theory, and then we derive from (12)
\be
\alpha \frac{d \tilde{\alpha}_{i}}{d\alpha} &=&
(\,-1+\frac{b^{(1)}_{i}}{b^{(1)}}\,)\, \tilde{\alpha}_{i}
-\sum_{j,k}\,\frac{b^{(1)}_{i,jk}}{b^{(1)}}
\,\tilde{\alpha}_{j}\, \tilde{\alpha}_{k}+\sum_{r=2}\,
(\frac{\alpha}{\pi})^{r-1}\,\tilde{b}^{(r)}_{i}(\tilde{\alpha})~,
\ee
where
$\tilde{b}^{(r)}_{i}(\tilde{\alpha})~,~r=2,\cdots$,
are power series of $\tilde{\alpha}_{i}$ and can be computed
from the $r$-th loop $\beta$ functions.
We then solve  the algebraic equations
\be
(\,-1+\frac{b^{(1)}_{i}}{b^{(1)}}\,)\, \rho_{i}
-\sum_{j,k}\frac{b^{(1)}_{i,jk}}{b^{(1)}}
\,\rho_{j}\, \rho_{k}&=&0~,
\ee
which give the fixed points of (13) at $\alpha=0$.
We  assume that the solutions $\rho_{i}$'s have the form
\be
\rho_{i}&=&0~\mbox{for}~ i=1,\cdots,N'~;~
\rho_{i} ~>0 ~\mbox{for}~i=N'+1,\cdots,N~,
\ee
and we regard $\tilde{\alpha}_{i}$ with $i \leq N'$
 as small
perturbations  to the
undisturbed system which is defined by setting
$\tilde{\alpha}_{i}$  with $i \leq N'$ equal to zero.
It is possible
\cite{zimmermann1} to verify at the one-loop level 
the existence of
the unique power series solutions 
\be
\tilde{\alpha}_{i}&=&\rho_{i}+\sum_{r=2}\rho^{(r)}_{i}\,
(\frac{\alpha}{\pi})^{r-1}~,~i=N'+1,\cdots,N~
\ee
of the reduction equations (13) to all orders  
in  the undisturbed system (as we will demonstrate it in our
$SO(10)$ model below).
These are RG invariant relations among couplings that keep formally
perturbative renormalizability of the undisturbed system.
So in the undisturbed system there is only {\em one independent}
coupling $\alpha$.

We emphasize that the more 
vanishing $\rho_i$'s a solution contains, the less is its 
predictive power in general. We therefore search for predictive 
solutions in a systematic fashion.

\subsection{Unperturbed system}
\noindent
{\bf (a) Model I}

We find that for  model I
there exist two independent
solutions,
$A$ and $B$, that have the most predictive 
power, where we have chosen the $SO(10)$ gauge coupling as
the primary coupling:
\be
\rho_{T} &= &\left\{
\begin{array}{ll} 163/60 &\simeq 2.717 \\
0 &  \end{array} \right. ~~,~~
\rho_{\phi} ~= ~\left\{
\begin{array}{ll} 5351/9180 &\simeq 0.583 \\
1589/2727 &\simeq 0.583 \end{array} \right. ~,\nn\\  
\rho_{S} &= &\left\{
\begin{array}{ll} 152335/51408 &\simeq 2.963 \\
850135/305424 &\simeq 2.783 \end{array} \right. ~~,~~
\rho_{A} ~= ~\left\{
\begin{array}{ll} 31373/22032 &\simeq 1.424 \\
186415/130896 &\simeq 1.424 \end{array} \right. ~,\nn\\ 
\rho_{HS}&= &\left\{
\begin{array}{ll} 7/81 & \simeq 0.086 \\
170/81 &\simeq 2.099 \end{array} \right. ~~,~~
\rho_{1NM} = \rho_{2NM} =\left\{
\begin{array}{ll} 191/204 &\simeq 0.936 \\
191/303 &\simeq 0.630 \end{array} \right. ~~,\nn\\
\rho_{3NM}&= &\left\{
\begin{array}{ll} 0 &  \\
191/303 &\simeq 0.630 \end{array} \right. ~
~~\mbox{for}~~\left\{\begin{array}{l} IA   \\
IB\end{array} \right.~.
\ee
Clearly, the solution B has less predictive power because
$\rho_T =0$. So, we consider below only the solution A,
in which the coupling $\alpha_{3NM}$ should be
regarded as a small perturbation because $\rho_{3NM}=0$.

Given this solution, we would like to show next (as promised)
that the expansion coefficients 
$\rho_{i}^{(r)}~,~i=T,\cdots, 2NM  $
can be uniquely computed in any 
finite order in perturbation theory.
To this end, we assume that $\rho_{i}^{(n)}$ with
$r \leq n-2$ are known, then 
insert the power series ansatz (16) for
$\rho_{i}^{(n)}~,~i \neq 3NM$ into the reduction equation (13)
and collect terms of $O(\alpha^{n-1})$. One finds easily that
\be
\sum_{j\neq 3NM} M_{i~j}(n)\,\rho_{j}^{(n)}
&=&~\mbox{known quantities by assumption}~,
~i\neq 3NM~,
\ee
where
\be
M(n) &=&
\left( \begin{array}{llll}
1141/30 - 7 n &0  & 0 & 0\\
0 & 283603/9180 - 7 n & 0 &
 21404/3825 \\
0 & 0 & 30467/612 - 7 n & 
152335/4284 \\
0 & 31373/1377 & 219611/27540 & 
909817/27540 - 7 n \\
 56/81 & 0 & 196/405 & 28/81\\
0 & 2865/136 & 0 & 0\\
0 & 2865/136 & 0 & 0 \\
\end{array} \right. \nn\\
& &~~~~~~~~~~~~~~~
\left. \begin{array}{lll}
 1467/100 & 0 & 0 \\
 0 & 5351/18360 & 5351/18360 \\
 152335/34272 & 0 & 0 \\
 31373/44064 & 0 & 0\\
 -791/810 - 7 n & 0 & 0 \\
  0 &
 573/68 - 7n & 191/24 \\
 0 &
 191/24 & 573/68 - 7 n\\
 \end{array} \right)~.
 \ee
So, if $\det M(n) \neq 0$, the coefficients
$\rho_{i}^{(n)}~,~i=T,\cdots,2NM$ can be uniquely calculated.
We in fact find
\be
\det M(n) &=&
-\frac{110920238635003554634381}{8522204882112000} + 
  n \frac{3608874567318092545318601}{25566614646336000}\nn\\
& &
+ n^2\frac{7571105122486669715209741}{8522204882112000} - 
  n^3\frac{391617250274453557751579}{284073496070400}\nn\\
& & + n^4\frac{598654192729460650727}{819127728000} - 
   n^5\frac{107001680791190563}{606761280}\nn\\
& &+ n^6 \frac{108620968687}{5508} - n^7 823543 \nn\\
&\neq& 0~~\mbox{for integer}~n~.
\ee
Therefore, there exists a unique power series solution
of the form (16)
for the solution IA.

\vspace{1cm}
\noindent
{\bf (b) Model II}

According to the principle of 
reduction of couplings, we search for
 most predictive solutions of (14).
Of these solutions, we consider only non-degenerate ones
with $\rho_T \neq 0$,
because they are more predictive. We find that there exist
three solutions, IIA, IIB and IIC,  and 
they contain
three vanishing $\rho$'s:
\be
\rho_{T} &= &\left\{
\begin{array}{ll} \frac{674137}{117840} &\simeq 5.7 \\
\frac{108764}{14225}& \simeq 7.7 \\
\frac{443}{60} &\simeq 7.4 \end{array} \right. ~,~
\rho_{GJ} = \left\{
\begin{array}{ll} \frac{674137}{2969568} &\simeq 0.2 \\
\frac{390223}{1433880} &\simeq 0.3\\
\frac{443}{1512} &\simeq 0.3 \end{array} \right. ~,~
\rho_{\phi} = \left\{
\begin{array}{ll} \frac{674137}{1060560} &\simeq 0.6 \\
\frac{300557}{512100} &\simeq 0.6 \\
\frac{443}{540} &\simeq 0.8 \end{array} \right.~, \nn\\
\rho_{S} &= &\left\{
\begin{array}{ll} 0 & \\
0 &  \\
\frac{16361}{4032} &\simeq 4.1 \end{array} \right. ~,~
\rho_{A} = \left\{
\begin{array}{ll} \frac{2356151}{5090688} &\simeq 0.5 \\
0 &  \\
\frac{4739}{1296} &\simeq 3. \end{array} \right. ~,~
\rho_{HS} = \left\{
\begin{array}{ll} \frac{1673861}{318168} &\simeq 5.3 \\
\frac{13811}{51210} &\simeq 0.3 \\
\frac{77}{81} &\simeq 1.0 \end{array} \right.~,\nn \\
\rho_{\Theta S} &= &\left\{
\begin{array}{ll} 0 &   \\
\frac{47260924}{4480875} &\simeq 10.5 \\
0 &   \end{array} \right. ~,~
\rho_{\overline{\Theta} S} = \left\{
\begin{array}{ll} \frac{ 673145}{371196} &\simeq 1.8 \\
\frac{ 1584242}{1493625} & \simeq 1.1\\
\frac{16937}{9450} &\simeq 1.8\end{array} \right. ~,~
\rho_{\Theta A} = \left\{
\begin{array}{ll} \frac{205226}{36825} &\simeq 5.6 \\
\frac{21623024}{4480875} &\simeq 4.8 \\
0 &  \end{array} \right. ~,\nn\\
\rho_{1NM} &= &\left\{
\begin{array}{ll} \frac{674137}{2969568} &\simeq 0.2 \\
\frac{53321}{159320} &\simeq 0.3 \\
\frac{443 }{1512} &\simeq 0.3 \end{array} \right. ~,~
\rho_{2NM} = \left\{
\begin{array}{ll} \frac{674137}{2969568} &\simeq 0.2 \\
\frac{53321}{159320} &\simeq 0.3 \\
\frac{443 }{1512} &\simeq 0.3\end{array} \right. ~,\nn\\
\rho_{3NM} & = &\left\{\begin{array}{ll}
0 &  \\
0 &   \\
0 &  \end{array} \right.~~\hspace{2cm} \mbox{for}
\left\{\begin{array}{ll}
\mbox{IIA} \\
\mbox{IIB}  \\
\mbox{IIC} \end{array} \right.~.
\ee 
Observe that certain $\rho$'s for solutions IIB and IIC 
are so large that the model cannot be treated in 
perturbation theory. The $\rho_T \simeq 5.7$ for solution IIA
could be within the regime of perturbation theory, but as we will
see in the next section, that value is  so large that the predicted
value of $M_t$ cannot be distinguished from its infrared value.
Therefore, this model does not yield a testable prediction
on the top quark mass.  

We presented in this section the negative result, too, in some detail
to emphasize that 
the existence of a consistent supersymmetric Gauge-Yukawa
unified model based on $SO(10)$ is a nontrivial matter,
as we have announced in the introduction.

\subsection{Small perturbations}
 The small 
 perturbations caused by nonvanishing $\tilde{\alpha}_{i}$
 with $i \leq N'$, defined in eq. (15) and
$\tilde{\alpha}_{3NM}$ in the
 case of solution IA,
enter in such a way that the reduced couplings, 
i.e., $\tilde{\alpha}_{i}$  with $i > N'$, 
become functions not only of
$\alpha$ but also of $\tilde{\alpha}_{i}$
 with $i \leq N'$.
It turned out \cite{kubo2} that, to investigate such partially
reduced systems, it is most convenient to work with the partial
differential equations, which  for solution IA  are
\be
\{~~\tilde{\beta}\,\frac{\partial}{\partial\alpha}
+\tilde{\beta_{3NM}}\,
\frac{\partial}{\partial\tilde{\alpha}_{3NM}}~~\}~
\tilde{\alpha}_{i}(\alpha,\tilde{\alpha})
&=&\tilde{\beta}_{i}(\alpha,\tilde{\alpha})~,~i\neq 3NM~,
\ee
where
\be
\tilde{\beta}_{i} &= &\frac{\beta_{i}}{\alpha^2}
-\frac{\beta}{\alpha^{2}}~\tilde{\alpha}_{i}
~, ~
\tilde{\beta}~=~\frac{\beta}{\alpha}~.\nn
\ee
These partial differential equations  are equivalent
to the reduction equations (13), and we
look for solutions of the form
\be
\tilde{\alpha}_{i}&=&\rho_{i}+
\sum_{r=2}\,(\frac{\alpha}{\pi})^{r-1}\,f^{(r)}_{i}
(\tilde{\alpha}_{3NM})~,~i\neq 3NM~,
\ee
where $ f^{(r)}_{i}(\tilde{\alpha}_{3NM})$ are supposed to be 
power series of 
$\tilde{\alpha}_{3NM}$. This particular type of solution
can be motivated by requiring that in the limit of vanishing
perturbations we obtain the undisturbed
solutions (16) \cite{kubo2}, i.e.,
$ f_{i}^{(r)}(0)=\rho_{i}^{(r)}~\mbox{for}~r \geq 2$.
Again it is possible to obtain  the sufficient conditions for
the uniqueness of $ f^{(r)}_{i}$ in terms of the lowest order
coefficients. The proof is similar to that for 
$\rho_{i}^{(r)}$'s.

We have computed these corrections up to and including terms of
$O(\tilde{\alpha}_{3NM}^2$):
\be
\tilde{\alpha}_T &=&
(\,163/60 - 0.108\cdots  \tilde{\alpha}_{3NM} + 
0.482 \cdots  \tilde{\alpha}_{3NM}^2+\cdots\,)
+\cdots~,\nn\\
\tilde{\alpha}_{\phi} &=&
(\,5351/9180 + 0.316\cdots  \tilde{\alpha}_{3NM} +
0.857\cdots  \tilde{\alpha}_{3NM}^2+\cdots\,)
+\cdots~,\nn\\
\tilde{\alpha}_{S} &=&
(\,152335/51408 + 0.573\cdots  \tilde{\alpha}_{3NM} 
+ 5.7504\cdots  \tilde{\alpha}_{3NM}^2+\cdots\,)
+\cdots~,\nn\\
\tilde{\alpha}_{A} &=&
(\,31373/22032 - 0.591\cdots  \tilde{\alpha}_{3NM} 
- 4.832\cdots  \tilde{\alpha}_{3NM}^2+\cdots\,)
+\cdots~,\\
\tilde{\alpha}_{HS} &=&
(7/81 - 0.00017\cdots  \tilde{\alpha}_{3NM} 
+ 0.056\cdots  \tilde{\alpha}_{3NM}^2+\cdots\,)
+\cdots~,\nn\\
\tilde{\alpha}_{1NM} &=&\tilde{\alpha}_{2NM}~=~
(\,191/204 - 4.473\cdots  \tilde{\alpha}_{3NM} 
+ 2.831\cdots  \tilde{\alpha}_{3NM}^2+\cdots\,)
+\cdots~,\nn
\ee
where $\cdots$ indicates higher order terms which
can be uniquely computed. 
In the partially reduced theory defined above,
we have two independent couplings,
$\alpha$ and $\alpha_{3NM} $ (along with the
Yukawa couplings $\alpha_{IJ}~,~ I,J\neq T$).

\begin{figure}
           \epsfxsize= 11 cm   
           \centerline{\epsffile{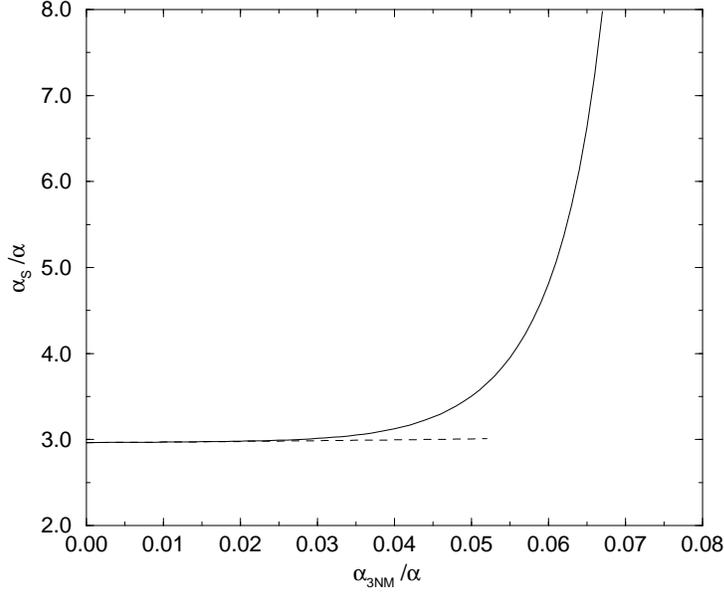}}
        \caption{$\tilde{\alpha}_{S}$ versus 
$\tilde{\alpha}_{3NM}$, where the dashed line is obtained from 
the analytic expression (24).}
        \label{fig:1}
        \end{figure}

At the one-loop level
eq. (24) defines a line parametrized by $\tilde{\alpha}_{3NM}$
in the $7$ dimensional space of couplings.
A numerical analysis shows  that this line blows up
in the direction of $\tilde{\alpha}_{S}$ at
a finite value of $\tilde{\alpha}_{3NM}$.
Fig. 1 shows $\tilde{\alpha}_{S}$ as a function
of  $\tilde{\alpha}_{3NM}$ (the dashed line is obtained
from the analytic expression (24)).
So if we require $\tilde{\alpha}_{S}$ to remain within
the perturbative regime (i.e., $g_S \lsim 2$,
which means $ \tilde{\alpha}_{S} \lsim 8$ because 
$\alpha_{\rm GUT} \sim 0.04$),
the $\tilde{\alpha}_{3NM}$ should be restricted to be below
$\sim 0.067$. As a consequence, the value of
$\tilde{\alpha}_{T}$ is also bounded. To see this, we plot
$\tilde{\alpha}_{T}$ as a function of $\tilde{\alpha}_{3NM}$
in fig. 2, from which we conclude that
\be
2.714 &\lsim& \tilde{\alpha}_{T} \lsim 2.736~.
\ee
This defines 
GYU boundary conditions holding 
at the unification scale $M_{\rm GUT}$ in addition to the 
group theoretic one,
$\alpha_{T}=\alpha_{t} ~=~\alpha_{b} ~=~\alpha_{\tau}$.
The value of $ \tilde{\alpha}_{T}$ is practically fixed so that in
the  following discussions we may assume that
$\tilde{\alpha}_{T}=163/60 \simeq 2.72$,
which is the unperturbed value.

  \begin{figure}
           \epsfxsize= 11 cm   
           \centerline{\epsffile{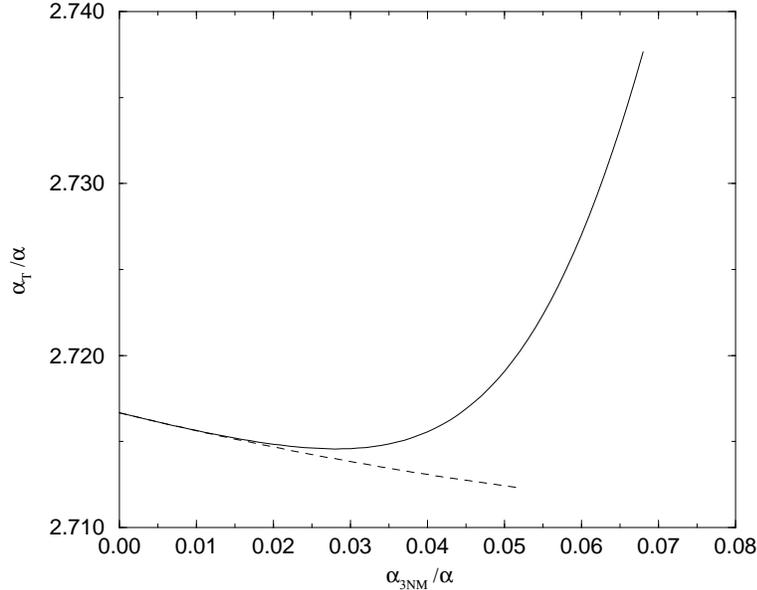}}
        \caption{$\tilde{\alpha}_{T}$ versus 
$\tilde{\alpha}_{3NM}$, where the dashed line is obtained from 
the analytic expression (24).}
        \label{fig:2}
        \end{figure}

\section{Predictions} 

 As pointed out,
 the GYU conditions (25) 
we have obtained above remain unaffected by
soft supersymmetry breaking terms,
because the $\beta$ functions are
not altered by these terms.
 To predict observable parameters from GYU,
we apply the 
renormalization group technique \cite{inoue1,so101}.

Just below $M_{\rm GUT}$ we would 
like to obtain the MSSM
while requiring that
all the superpartners are decoupled below 
the supersymmetry breaking scale $M_{\rm SUSY}$.
 To simplify our numerical analysis  we assume a
unique  threshold $M_{\rm SUSY}$ for all the  superpartners.
Then the SM should  be spontaneously broken down to
$SU(3)_{C}\times U(1)_{EM}$ due to the vacuum expectation
value of the scalar component of $H_\alpha$.
We also assume that the low energy theory which satisfies the
requirement above can be obtained  by arranging
soft supersymmetry breaking terms and
the mass parameters in the superpotential (3)  
in an appropriate fashion.

We shall examine numerically 
the evolution of the gauge
and Yukawa couplings including the two-loop
 effects, according to their renormalization
group equations \footnote{We  take into account
the threshold effects by the step function approximation
of the $\beta$ functions.}.
The translation of the value of a Yukawa coupling
into the corresponding mass value follows according to
$m_i ~= ~g_i(\mu)\,v(\mu)/\sqrt{2}~,~i=t,b,\tau$,
where $m_i(\mu)$'s are the running masses and we use
$v(M_Z)=246.22~\mbox{GeV}$.
The pole mass $M_i$ can be calculated from the
running one, and  for the top mass, we use \cite{so101}
\be
M_{t} &=&m_{t}(M_t)\,[\,1+
\frac{4}{3}\frac{\alpha_3(M_t)}{\pi}+
10.95\,(\frac{\alpha_3(M_t)}{\pi})^2\,]~,
\ee
where  we compute $v(M_t)$ from $v(M_Z)$.
As for the tau and bottom masses, we assume that
$m_{\tau}(\mu)$ and $m_b(\mu)$ for $\mu < M_Z$
satisfy the evolution equation governed by
the $SU(3)_{\rm C}\times U(1)_{\rm EM}$ theory
with five flavors and use \cite{so101}
\be
M_{b}&=&m_b(M_b)\,[\,1+
\frac{4}{3}\frac{\alpha_{3(5{\rm f})}(M_b)}{\pi}+
12.4\,(\frac{\alpha_{3(5{\rm f})}(M_b)}{\pi})^2\,]~,\nn\\
M_{\tau}&=&m_{\tau}(M_{\tau})\,[\,1+
\frac{\alpha_{\rm EM (5f)}(M_{\tau})}{\pi}\,]~,
\ee
where the couplings with five flavors
$\alpha_{3(5{\rm f})}$ and $\alpha_{\rm EM (5f)}$
are related to $\alpha_{3}$ and $\alpha_{\rm EM}$ by
\be
\alpha_{3(5{\rm f})}^{-1}(M_Z) &= &\alpha_{3}^{-1}(M_Z)
-\frac{1}{3\pi}\frac{M_t}{M_Z} ~,\\
\alpha_{\rm EM (5f)}^{-1}(M_Z) &= & \alpha_{\rm EM}^{-1}(M_Z)-
\frac{8}{9\pi}\frac{M_t}{M_Z}~.
\ee
The corrections in eq. (27) are  the SM ones, and 
in general one should
add  the MSSM corrections too. They could be even large,
especially for $M_b ~\sim \pm O (20-30 \%)$, in the case of
universal soft supersymmetry breaking terms,  while they can be kept
small  if these terms are not universal \cite{so102,so103}. As we will
see below, our prediction for $m_b(M_b)$ without the MSSM
corrections fit to the experimental value so that
the model favors the non-universal soft supersymmetry breaking terms.

Regarding now
\be
M_{\tau} &=&1.777 ~\mbox{GeV}~,~M_Z=91.188 ~\mbox{GeV}~,\nn\\
\alpha_{\rm EM}^{-1}(M_{Z}) & =&127.9
+\frac{8}{9\pi}\,\log\frac{M_t}{M_Z} ~,\nn\\
\sin^{2} \theta_{\rm W}(M_{Z})&=&0.2319
-3.03\times 10^{-5}T-8.4\times 10^{-8}T^2~,\\
T &= &M_t /[\mbox{GeV}] -165~,\nn
\ee
as given \cite{pdg,pokorski1}, we find
\be
m_{\tau}(M_{Z}) &=&1.746~\mbox{GeV}~,~
\alpha_{\tau}(M_Z)~=~\frac{g_{\tau}^{2}}{4\pi}
=8.005\times 10^{-6}~,\nn
\ee
which, together with $\alpha_{\rm EM}$ and 
$\sin^{2}\theta_{\rm W}$ given in (30),  we use as the input for the 
RG evolution.
In fig. 3, 4 and 5, we show the predictions of model IA on
$M_t$, $m_b (M_b)$ and $\alpha_3 (M_Z)$, respectively.
Note that the mass values are  before the MSSM
corrections are taken into account. Since $ m_b(M_b)$ agrees with the
experimental value ($4.1-4.5$) GeV \cite{pdg}
as we can see from fig. 4, these corrections should
be rather small, implying that our model favors the non-universal soft
supersymmetry breaking terms \cite{so103}.

 \begin{figure}
           \epsfxsize= 11 cm   
           \centerline{\epsffile{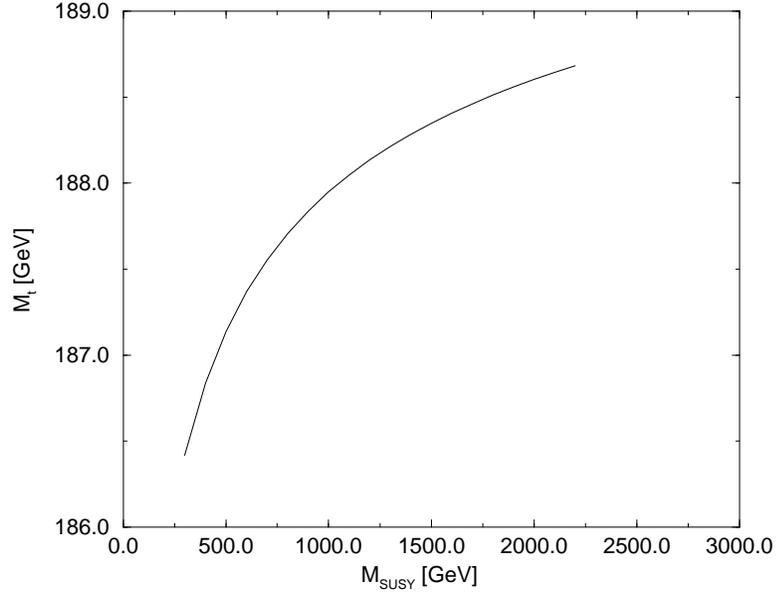}}
        \caption{$M_t$ prediction versus 
$M_{\rm SUSY}$ for $\tilde{\alpha}_{T}=2.717$.}
        \label{fig:3}
        \end{figure}
 \begin{figure}
           \epsfxsize= 11 cm   
           \centerline{\epsffile{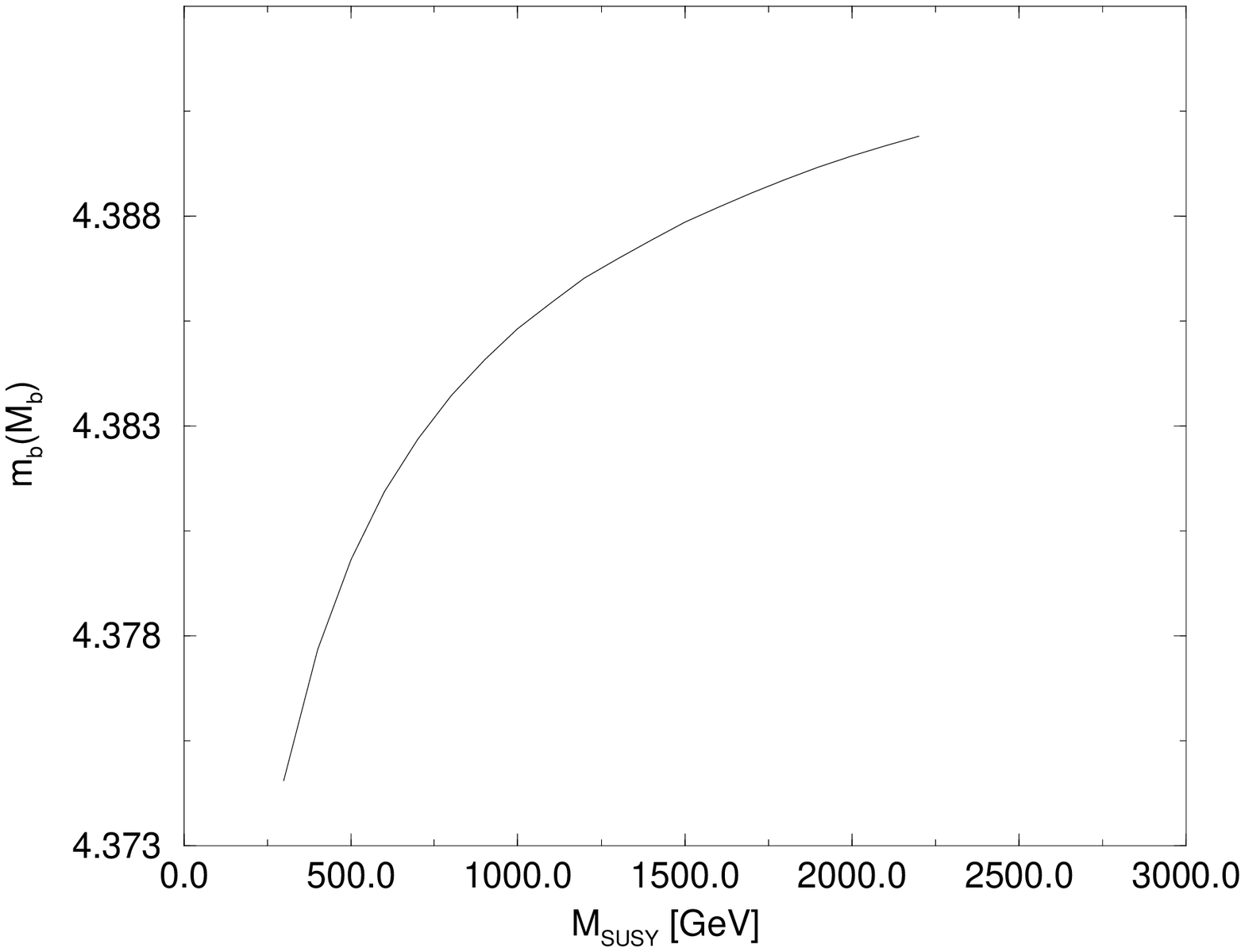}}
        \caption{$m_{b}(M_b)$ prediction versus 
$M_{\rm SUSY}$ for $\tilde{\alpha}_{T}=2.717$.}
        \label{fig:4}
        \end{figure}
 
\begin{figure}
           \epsfxsize= 11 cm   
           \centerline{\epsffile{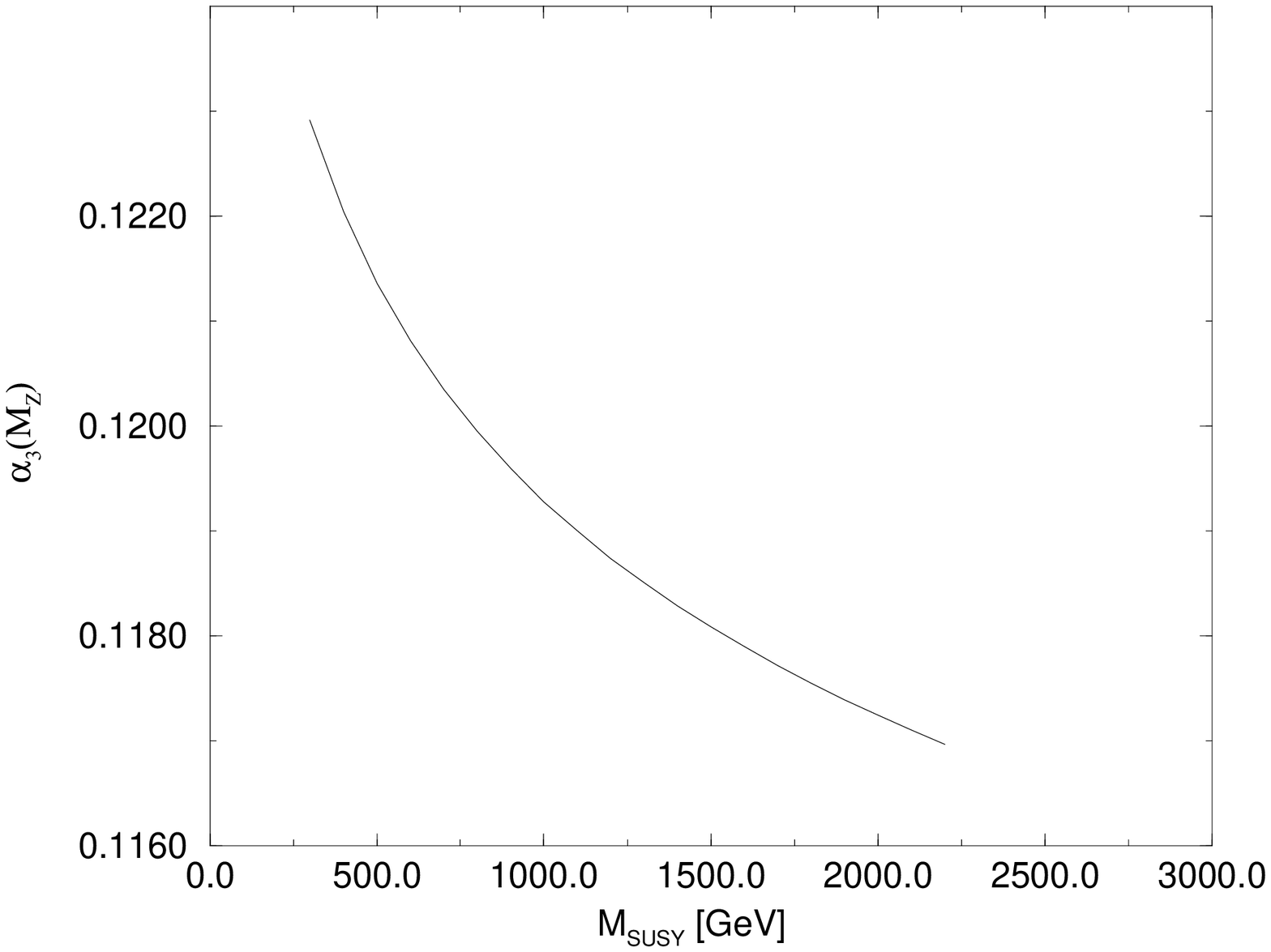}}
        \caption{$\alpha_{3}(M_Z)$  prediction versus 
$M_{\rm SUSY}$ for $\tilde{\alpha}_{T}=2.717$.}
        \label{fig:5}
        \end{figure}

Fig. 6 shows that $M_t$ is relatively insensitive 
to a change in $\tilde{\alpha}_T$.
The reason is that the predicted values for $M_t$ are not very much far
from its infrared value \cite{infra}, which we define as the value
for $\tilde{\alpha}_T=6$ and is shown in fig. 7.
By comparing figs. 6 and 7, we see that the predicted values for
$M_t$ lie a few GeV below the infrared values.
So, if the experimental uncertainty can be reduced to
less than that order, there will be a chance
to test the Gauge-Yukawa unified model we have proposed here.
Note that the present experimental value of $M_t$ is
$(180\pm 12)$ [GeV] \cite{top}.

 \begin{figure}
           \epsfxsize= 9 cm   
           \centerline{\epsffile{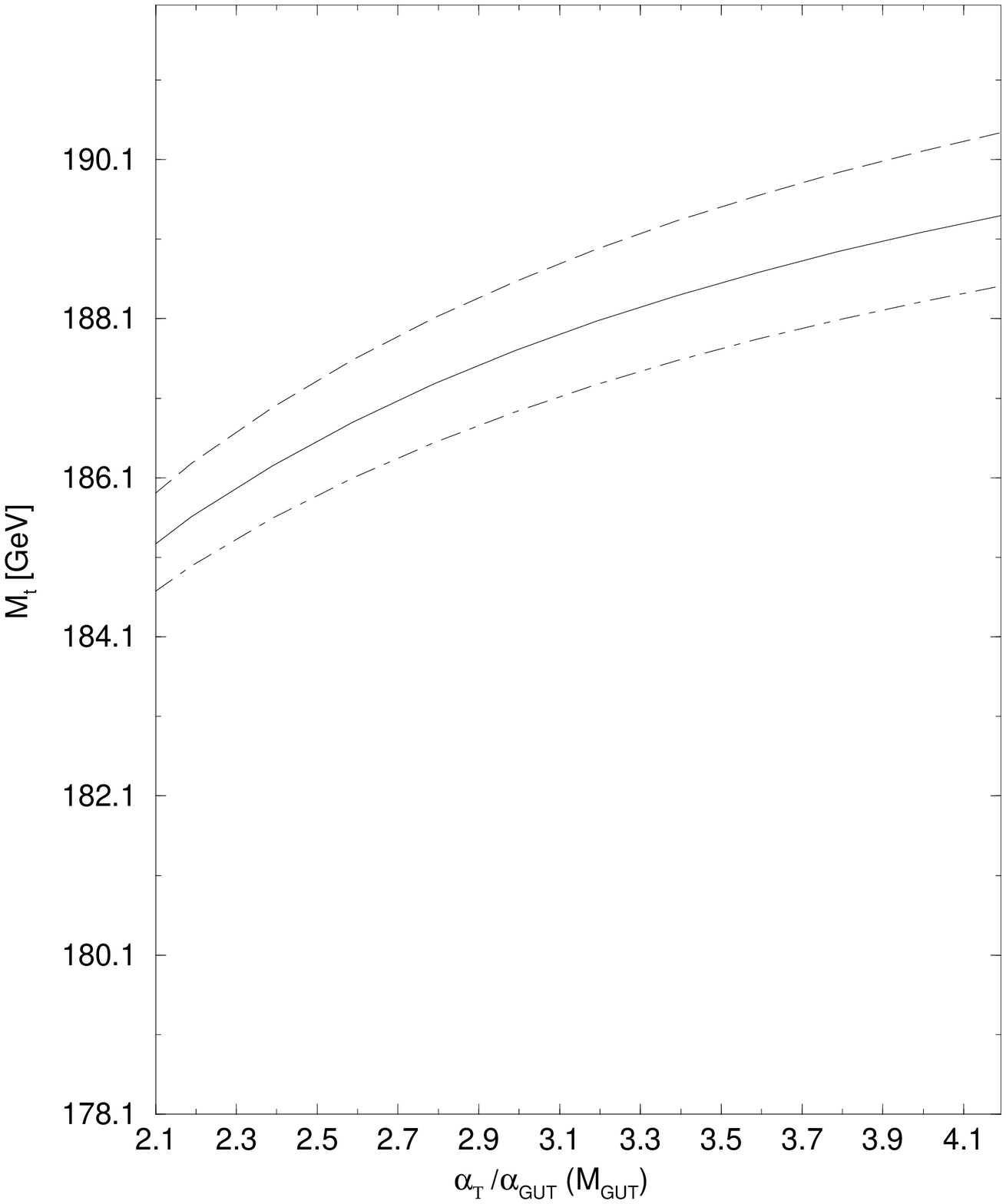}}
        \caption{$M_t$ versus 
$\tilde{\alpha}_{T}$ with $M_{\rm SUSY}=300$ GeV
(dod-dashed line), $500$ GeV (solid line)
and $1$ TeV (dashed line).}
        \label{fig:6}
       \end{figure}

 \begin{figure}
           \epsfxsize= 11 cm   
           \centerline{\epsffile{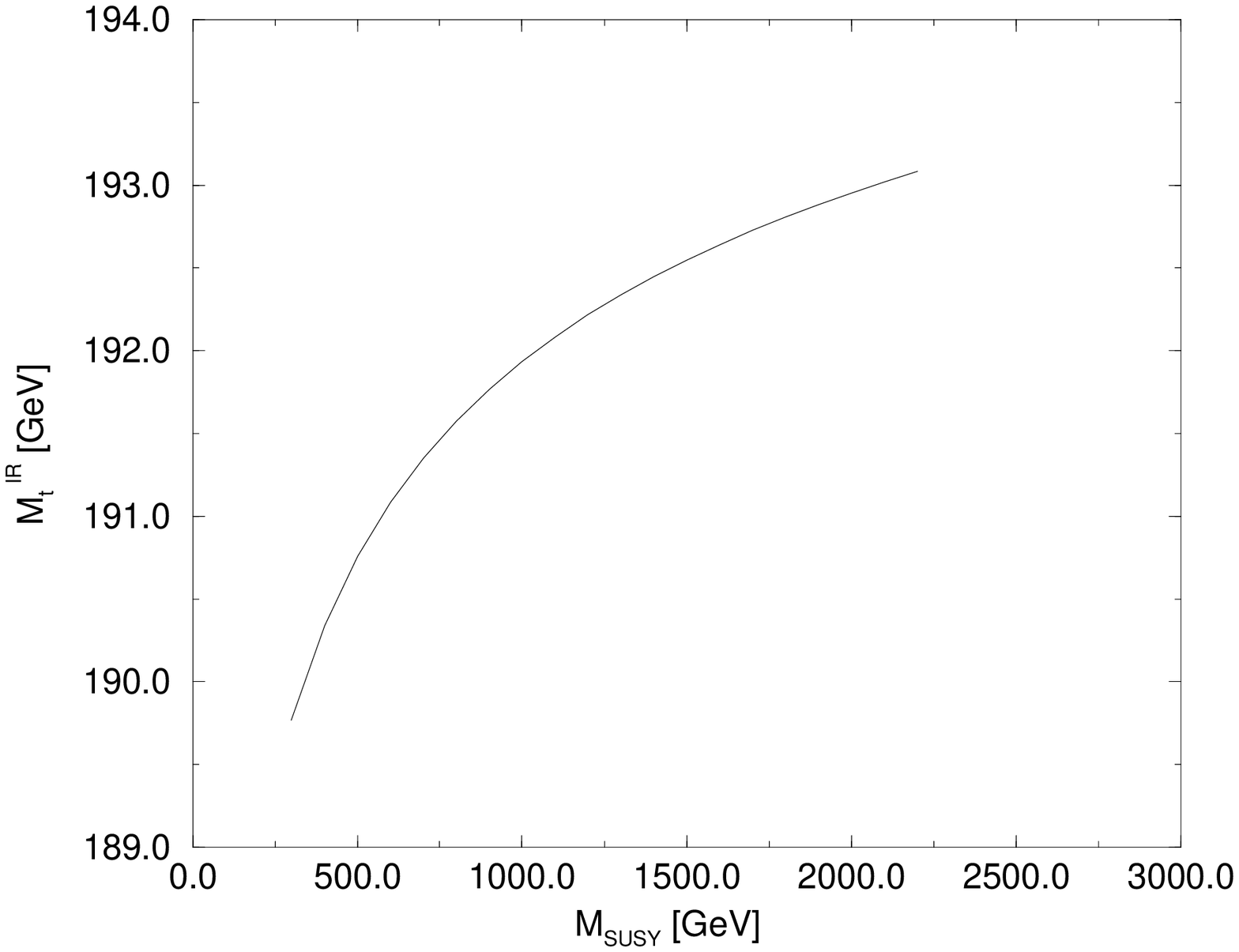}}
        \caption{The infrared value of $M_t$ versus 
$M_{\rm SUSY}$ ($\tilde{\alpha}_{T}=6$).}
        \label{fig:7}
        \end{figure}

Finally, we would like to comment on the difference
of the prediction on $M_t$ with and without GYU.
As pointed out, the $M_t$  prediction without GYU
follows from the requirement of the consistent
bottom-tau hierarchy. Under the same assumption made for the 
RG analysis above, we wish to find the allowed values
of $\tilde{\alpha}_T$. We find that the requirement
of $ 4.2~\mbox{GeV} \lsim 
m_b(M_b) \lsim 4.5 ~\mbox{GeV}$ with 
$M_{\rm SUSY}=500$ GeV fixed, for instance, 
implies $   1.6 \lsim 
\tilde{\alpha}_T \lsim 7.0$, and consequently, 
$ 182.9~\mbox{GeV} \lsim 
M_t \lsim 191.2 ~\mbox{GeV} $. This
should be compared with the GYU 
prediction,
\be 
m_b(M_b) &=&4.38 ~\mbox{GeV}~,~
M_t=187.1 ~\mbox{GeV}~,
\ee which
is fixed under the same 
assumption.

\section{Comparison with the infrared-fixed-point
approach and discussions} 

The infrared-fixed-point approach \cite{ross1} 
is based on the assumption
that infrared fixed points found in first order 
in perturbation theory 
persist in higher orders \footnote{In the case of the SM, the
infrared fixed point
approach loses its meaning at the two 
loop-level \cite{zimmermann3}.}
and that the ratio of the compactification scale $\Lambda_{\rm C}$
(or the Planck scale $M_{\rm P}$) to
$M_{\rm GUT}$ is large enough for 
the ratio of the top Yukawa coupling  to the gauge coupling
to come very close to its infrared value when running
from $\Lambda_{\rm C}$  down to $M_{\rm GUT}$.
Since this approach looks similar to ours at first sight,
we would like to examine within the  framework of
our $SO(10)$ model whether and how much this 
picture of infrared-fixed-point behavior is
realized. 

To this end, let us first re-capitulate the argument
of ref. \cite{ross1} and recall  that \footnote{
Our normalization of the generators of $SO(10)$ differs by a factor
of $\sqrt{2}$.  Our choice corresponds to that of the usual
$SU(5)$ GUTs.}
\be
\alpha(\Lambda_{\rm C}) &=& 
\frac{\alpha(M_{\rm GUT})}{
1-(7/2\pi)\alpha(M_{\rm GUT})
\ln (\Lambda_{\rm C}/M_{\rm GUT})}~
\ee
for one-loop order.
For $\alpha(M_{\rm GUT})= 0.04$ and 
$ \ln (\Lambda_{\rm C}/M_{\rm GUT})=5$,
we obtain
$ \alpha(\Lambda_{\rm C}) \simeq  0.051~$.
Assuming now that $\alpha_i$'s with $i\ne T$ are
negligibly small compared with $\alpha_T$,  
we derive from the reduction equation (13)
\be
\alpha\frac{\tilde{\alpha}_T}{d \alpha} &=&
\tilde{\alpha}_T (2 \tilde{\alpha}_T-\frac{11}{2})~,
\ee
with the solution \cite{ross1,kubo1}
\be
\tilde{\alpha}_T (M_{\rm GUT}) &=&
\frac{1}{4/11+
(\kappa_{T}^{-1}-4/11)[\alpha(M_{\rm GUT})
/\alpha(\Lambda_{\rm C})]^{11/2}}~,
\ee
where $\kappa_T$ is the value of $\tilde{\alpha}_T$
at $ \Lambda_{\rm C}$.
The point in the infrared-fixed-point approach
 is that, since 
 $[\alpha(M_{\rm GUT}) /\alpha(\Lambda_{\rm C})]^{11/2}$ 
is small ($\sim 0.25$), 
the ''low-energy'' value, $\tilde{\alpha}_T (M_{\rm GUT})$, is 
insensitive against $\kappa_T$ and must be close to
its infrared value $11/4 =2.75$.
One indeed finds that one may vary $\kappa_T$ from $ 1.9$
to $4.3$ to keep
 $ [\tilde{\alpha}_T (M_{\rm GUT})/(11/4)-1] \lsim 0.1$. It however
should be emphasized that there is no principle to fix $\kappa_T$
in the infrared-fixed-point approach. 
For $\kappa_T=1$, which could be realized with the
same probability as for $\kappa_T=2$,
we obtain  $\tilde{a}_T (M_{\rm GUT})\simeq 1.91$,
which is only $70 \%$  of the infrared value.

 \begin{figure}
           \epsfxsize= 11 cm   
           \centerline{\epsffile{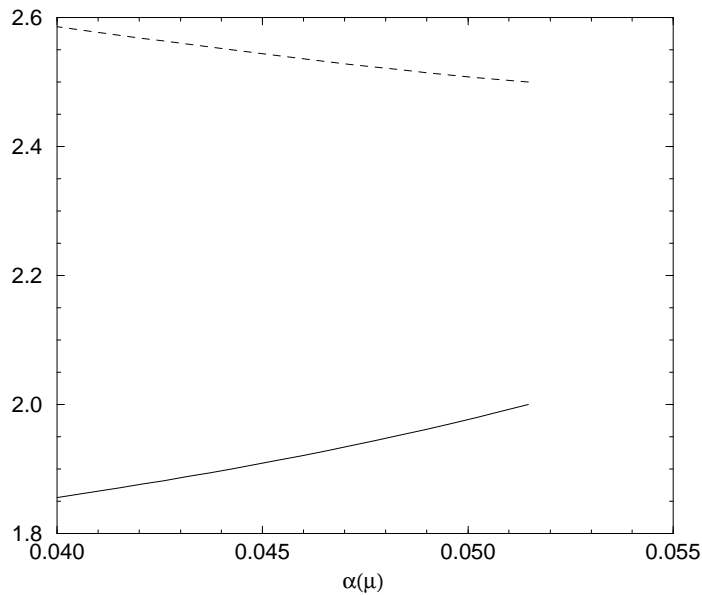}}
        \caption{$\tilde{\alpha}_{T}
(M_{\rm GUT})$ (solid line) and $\tilde{\alpha}_{HS}
(M_{\rm GUT})$ (dashed line)
versus $\alpha(\mu)$ for $\kappa_T=2$ and
$\kappa_{HS}=2.5$.}
        \label{fig:8}
        \end{figure}

The more serious problem is  the negligibility of
other couplings compared to $\alpha_T$.
 Since there exists no reason
why the neglected couplings have to be small,
they could be large and hence comparable to $\alpha_T$,
thereby changing the infrared structure  very much.
In fig. 8, we plot $\tilde{\alpha}_T (M_{\rm GUT})$
as a function of $\alpha(\mu)$ with $
0.04 \leq \alpha(\mu) \leq 0.051$, where we have chosen
$\kappa_T=2$ and $\kappa_{HS}=2.5$ while neglecting
the other couplings in the evolution. 
As we can see from fig. 8, 
the $\tilde{\alpha}_T$ does not approach to $2.75$
as $\alpha$ goes to $\alpha_{\rm GUT}=0.04$ from $0.051$,
rather to another fixed point $1.67$.
(If $\kappa_{HS}$ would be zero, the
$\tilde{\alpha}_T(M_{\rm GUT})$  would become
$2.51$.)

The observation of ref. \cite{ross1} 
that, in spite of the small difference between 
$M_{\rm GUT}$ and $\Lambda_{\rm C}$, the Yukawa
couplings tend to converge to their
fixed points very fast thanks to 
large anomalous dimensions of the matter superfields, is 
generally correct in one-loop order.
However,  the 
infrared-fixed-point approach may not always have predictive power,
as we have explicitly seen above in our concrete model.
This is not a specific situation of the present $SO(10)$ model,
because the factor 
$[\alpha(M_{\rm GUT}) 
/\alpha(\Lambda_{\rm C})]^{11/2}\simeq 0.25 $ 
is small enough according to the discussion
of ref. \cite{ross1}.
If one insists from the beginning to choose
 the one-loop infrared fixed point which is most
predictive, the lowest order prediction is exactly the same as 
that of the reduction of couplings.
The difference appears in the next order, because
the reduction solution, except for the
lowest order, is not a fixed point solution in
general.

If the low-energy prediction for either approach
is viable as experienced in some cases including the
one discussed in this paper, it
might indicate some unknown  non-perturbative
mechanism of unification of
couplings such as
the dynamical unification of couplings which we suggested
in introduction. In either case, 
the non-perturbative investigation on
asymptotically-nonfree, non-abelian gauge theories
will be an important issue in future works.

\vspace{1cm}
\noindent
{\bf Acknowledgment}

We thank R.N. Mohapatra for a useful correspondence, and
one of us (G.Z.) would like to thank P. Minkowski
for stimulating discussions.

\end{document}